\documentstyle{article}

\newcommand{\mathbf}{\bf}

\begin{document}

\begin{center}
{\huge\bf A New Uncertainty Relation}
\end{center}

\vspace{1cm}
\begin{center}
{\large\bf
F.GHABOUSSI}\\
\end{center}

\begin{center}
\begin{minipage}{8cm}
Department of Physics, University of Konstanz\\
P.O. Box 5560, D 78434 Konstanz, Germany\\
E-mail: ghabousi@kaluza.physik.uni-konstanz.de
\end{minipage}
\end{center}

\vspace{1cm}

\begin{center}
{\large{\bf Abstract}}
\end{center}

\begin{center}
\begin{minipage}{12cm}
We formulate according to the quantum mechanical uncertainty  
relation a new uncertainty relation
$\Delta \breve{A} \cdot \Delta l \sim  
\displaystyle{\frac{\hbar}{e}}$ where $\breve{A}$ and $\Delta l \geq  
l_B$ are the electromagnetic {\it pure} gauge potential, the  
position uncertainty and the magnetic length, respectively. Then, we  
show that the observed potential drops on the edge of QHE samples  
are varifications of this uncertainty relation, where the quantum  
potential drop of the relevant component of potential can be  
considered as its quantum uncertainty $\Delta \breve{A}$.

\end{minipage}
\end{center}

\newpage
First let us mention some physical background of the new  
uncerytainty relation, although its mathematical derivation from  
flux quantization and also its experimental varifications are very  
general and entirely independent of these backgrounds.

One can show that the microscopic theory of QHE \cite{all} should  
be given by the canonical quantization of a semi-classical theory of  
the "classical" Hall-effect CHE \cite{mein}.
The action functional for this is the semi-classical  
Schroedinger-Chern-Simons action for a 2-D non-interacting carrier  
system with the usual minimal electromagnetic coupling on a   
2+1-dimensional manifold $M = \Sigma\times\mathbf R$ with spatial  
boundary. It can be shown also that the constraints of the theory  
forces the coupled electromagnetic potential to be an almost pure  
gauge potential, i. e. with an almost vanishing field strength and  
they forces also the potential to exist only on the edge of sample  
\cite{mein}. Thus, according to our model we have to do in IQHE case  
with an almost pure "edge" gauge potential \cite {almost}.  
Accordingly, in view of Ohm's equations the edge currents are the  
prefered currents under these constraints of the theory in our model  
\cite{mein}.

The classical theory [2] requires that edge current of electrons  
should flow exactly on the edge of QHE-sample, i.e. with a zero  
distance to the edge of sample. However quantum theory forbids, in  
view of uncertainty relations, statements about zero distances in  
case of quantum systems or particles. Such a localization will  
require an infinite amount of momentum change. Therefore, edge  
currents are defined {\it quantum theoretically} as currents which  
flow within a distance of $\Delta q = l_B$ from the edge of sample  
\cite{kk}.

Moreover, as like as in quantum mechanics where the position  
uncertainty $\Delta q$ is correlted with a momentum uncertainty  
$\Delta p$. We will show that in the same manner there is an  
uncertainty for the value of electromagnetic {\it pure} gauge  
potential $\Delta \breve{A}$ for charged systems in strong magnetic  
fields according to the related uncertainty relations. Recall that,  
from the flux quantization $\oint e \breve{A}_m dx^m \propto \hbar$,  
a pure {\it quantized} gauge potential $\hat{\breve{A}}_m$ is  
comparable with $\hat{P}_m$ in view of its definition by the  
covariant derivative in QED $D_m \Psi \propto (\hat{P}_m - e  
\breve{A}_m) \Psi := 0$ and is of the same dimension $L^{-1}$ ( see  
also below).

Recall also, that from the local vanishing of the electronic  
current $j_m = ne V_m = \Psi^* (\hat{P}_m - e \breve{A}_m) \Psi$  
{\it on the integration path} of flux quantization $\oint  
\breve{A}_m dx^m$ \cite{current}, one concludes that $[ \hat{V}_m \  
, \ \hat{x}^m ] = 0$ {\it (on this path}) which implies $[\hat{P}_m  
\ , \ \hat{x}^m ] = [ e \hat{\breve{A}}_m \ , \ \hat{x}^m ]$, i. e.  
$[ e \hat{\breve{A}}_m s , \ \hat{x}^m ] = -i \hbar$ (on this path).  
From general quantum mechanics we know that the generators of the  
Heisenberg algebra, i. e. $[ \hat{f} \ , \ \hat{g} ] = -i \hbar$  
fulfil the uncertainty relation $\Delta f \cdot \Delta g \sim \hbar$  
\cite{qml}. Therefore, there is an uncertainty relation $e \Delta  
A_m \cdot \Delta x^m \sim \hbar$.

Moreover it is well known that just in the cases under  
consideration, namely in QHE or in flux quantization, the coordinate  
operators of the centre of cyclotron motion as well as the relative  
coordinates around the centre for electrons under magnetic fields  
are non-commuting, e. g. $[\hat{x^m} \ , \ \hat{x^n}] = -i l_B ^2  
\epsilon^{mn}$ for relative coordinates. This is an interesting  
example of the non-commutative geometry in quantum theory. Now, the  
commutator $[ \hat{A_m} \ , \ \hat{x^m}]$ is proportional to this  
commutator in the usual Landau gauge $A_m = B x^n \epsilon_{mn}$.  
Therefore, in view of this proportionality one have indeed $[  
\hat{A_m} \ , \ \hat{x^m}] = B \epsilon_{mn} [\hat{x^m} \ , \  
\hat{x^n}] = -i l_B ^2 \cdot B = -i \displaystyle{\frac{\hbar}{e}}$  
for $C = 1$, which is equivalent to the uncertainty relation $e  
\Delta A_m \cdot \Delta x^m \sim \hbar (no \  summation)$  
\cite{qml}.

\medskip
It should be mentioned also that the usual argument, that the  
electromagnetic potential $A_m$ is a function of $x^m$ and therefore  
their operators must commute with each other, does not apply to the  
case under consideration where we have to do with large {\it  
constant} magnetic field $B$ and edge currents as in QHE and flux  
quantization:
$\oint e A_m dx^m = \int \int e B \epsilon_{mn} dx^m dx^n = \phi  
\hbar = N h$.

Here $A_m$ is not a function of $x^m$, but it is given {\it on the  
closed path} of integration (the "circle") as an electromagnetic  
pure gauge potential $\breve{A}_m := \partial_m \phi$ with $F_{mn}  
(\breve{A}_m)_{(on)} = 0$ \cite{current}. Or it is given {\it  
within} the surface surrounded by the mentioned path as $A_m = B  
\cdot x^n \epsilon_{mn}$ (Landau gauge) where $F_{mn} (A_m) =  
B_{(in)} \epsilon_{mn}$ is constant, which is in accordance with the  
above discussed Heisenberg algebras $[x^m , x^n] \propto [A_m ,  
x^m]$.

As an example of the pure gauge potential one may consider  
$\breve{A}_l = \displaystyle{\frac{\partial \phi}{\partial l}} =  
\displaystyle{\frac{1}{R}}$, where $R$ is the radius of the  
integration circle in $\oint_{(on)} \breve{A}_m dx^m = \oint_{(on)}  
\breve{A}_l dl$ and $ 0 \leq l \leq 2N \pi R$ is the variable on the  
circle. Thereby, the function $\phi = \displaystyle{\frac{l}{R}}$  
should be a phase angle in order that the flux $\oint_{(on)} e  
\breve{A}_l dl = \int \int_{(in)} e F_{mn} dx^m \wedge dx^n = \hbar  
\phi$ is quantized according to the winding number of the  
integration path, i. e. by $\phi = 2\pi N \ , \ N \in {\mathbf Z}$.  
Recall again, that $F_{mn}$ is the electromagnetic field strength  
which is present within the surface surrounded by the integration  
path, whereas the pure gauge potential $\breve{A}_l$ has to be  
considered on this path, which becomes a ring of the width $\Delta x  
\sim l_B$ in view of quantization.

\medskip
We will prove that the recent results on the potential drops across  
IQHE samples near the edges \cite{DK} and \cite{font} follow the  
universal uncertainty relations of quantum electrodynamics, in  
accordance with the universality
of QHE \cite{more}.

We give here for the existence of such an uncertainty relation a  
proof according to the general quantum algebraic structure  
("operator structure" or "commutator structure") of quantum  
mechanics which should be fulfilld in any regular quantum theory. It  
is a result of canonical quantization structure which should be  
applicable in any regular quantum theory. This is so, because all  
other quantization formalisms should be equivalent to the canonical  
one.

\medskip
We will show that, indeed for the true phase space-variables of the  
flux quantization system, i. e. for electromagnetic systems under  
strong magnetion fields where the flux quantization take place, the  
commutator of the related operators is non-zero and so there exist  
an uncertainty relation which is varified experimentally by the  
potential drops experiments in QHE. The key point here is the choise  
of correct phase space for the electrodynamical system under  
consideration which has to be quantized in order to describe the  
flux quantization. In view of the fact that flux quantization is an  
experimental fact, the question is how to describe this fact  
theoretically. In other words, we should look for the quantization  
of a phase space which describes the flux quantization or the  
electromagnetic quantization under strong magnetic fields.
Such a canonical quantization has the advantage to give a general  
theoretical model for flux quantization and to introduce a new  
uncertainty relation, which  explain some experimental results in  
QHE.

\medskip
Nevertheless, I describe here the very general canonical  
quantization of an electromagnetic system for flux quantization  
which is in accordance with the above discussions.

It is well known that the quantization of Maxwell's action  
functional can not explain the flux quantization, thus we have to  
look for the quantization of another action functional which can  
desribe it.
However, although the flux quantization is different than the  
quantization in Maxwell's electrodynamics, the general canonical  
quantization used here apply to both of them.

The point of departure is the flux quantization relation for  
electromagnetic systems under strong magnetic fields:

\begin{equation}
\int \int e F_{mn} dx^m \wedge dx^n = \oint e \breve{A}_m dx^m =  
\phi \hbar
\end{equation}
\label{one}

Now, because this quantization is varified experimentally,  
therefore it should be describable "theoretically" as the canonical  
quantization of the classical action functional:

\begin{equation}
S_{(Cl)} = \oint e \breve{A}_m dx^m = \int \int e d\breve{A}_m  
\wedge dx^m = \int \int e F_{mn} dx^m \wedge dx^n ,
\end{equation}
\label{two}

which has to be quantized to describe the flux quantization  
according to (1):

\begin{equation}
S_{(Cl)} \rightarrow S_{(Q)} = \int \int e F_{mn} dx^m \wedge dx^n   
= \phi \hbar
\end{equation}
\label{three}

To quantize any action functional $S$, i. e. to quantize the  
variables involved in the phase space of a system represented by the  
action functional $S$ {\it in the canonical way}, one should  
compare such an action functional $S$ with the general canonical  
action functional:

\begin{equation}
S_{(canon)} = \int \int dP_m \wedge dx^m - \int \int dH \wedge dt
\end{equation}
\label{four}

of the same dimension.

From the point of view of symplectic structure and of the rigorous  
methode of geometric quantization \cite{wood}, the first term in  
action functional is enough to postulate the canonical quantization  
by

$\int \int dP_m \wedge dx^m = {\mathbf Z} \hbar$ which is  
equivalent to the commutator postulate $[ \hat{P}_m \ , \ \hat{x}^n  
] = -i \hbar \delta_m ^n$. However, taking also the second term in  
(4) into account, because in our $S_{(Cl)}$ there is no second term  
which contains explicitey the time integration, we have to compare  
our $\int \int e d \breve{A}_m \wedge dx^m$ with the canonical $\int  
\int dP_m \wedge dx^m$ term in order to identify the true variables  
of the phase space of our system \cite{comp}.

This canonical comparision shows that the phase space of our  
electrodynamical system, which is represented by the action $S_{Cl}$  
of (2) has the set ${\{ e\breve{A}_m, x^m}\}$ of canonical  
conjugate variables.

\bigskip
Then, the true globally Hamiltonian vector fields of our system  
with the symplectic 2-form

$\omega = e d\breve{A}_m \wedge dx^m$ are given by \cite {wood},  
\cite{erk}:

\begin{equation}
X_{\breve{A}_m} = \displaystyle{\frac{\partial}{\partial x^m}}  
\;\;\; , \;\;\; X_{x^m} = - \displaystyle{\frac{\partial}{\partial  
\breve{A}_m}}
\end{equation}
\label{five}

Moreover, the quantum operators on the quantized phase space of  
this system should be proportional to these vector fields by a  
complerx factor, i. e. usually by $(-i \hbar )$ or by $\hat{A} = -i  
\hbar \displaystyle{\frac{\partial}{\partial x^m}}$ and  $\hat{x} =  
i \displaystyle{\frac{\partial}{\partial \breve{A}_m}}$ \cite{clq}.

On the other hand, the actual phase space of motion of system  
should be polarized in the sense that the classical action and also  
the wave function should be functions of only half of the variables  
of the original phase space \cite{wood}. This means that in general  
$\Psi$ is either $\Psi ( P_i , t)$ or $\Psi (x^i , t)$. Then, the  
half of quantum operators which are related to the variables in  
$\Psi$ act on $\Psi$ just by the multiplication with these variables  
and the second half of quantum operators act on it by the action of  
quantum operators discussed above. In other words, as it is well  
known, for example in the $\Psi ( P_i , t)$ representation the  
acting operators are given by $\hat{x}^i = - i\hbar  
X_{x^i}^{(canon)} = i \hbar \displaystyle{\frac{\partial}{\partial  
P_i}}$ and $\hat{P}_i = P_i$, which result in the correct  
commutators: $[\hat{P}_i \ , \ \hat{x}^j ] = -i \hbar \delta_i ^j$.

In our case, where in view of the neccessary polarization the wave  
function of our ${\{\breve{A}_m , x^m}\}$ system is either in $\Psi  
( \breve{A}_m , t)$ or in $\Psi ( x^m , t)$ representation, the  
quantum operators are given either by ${\{ \hat{\breve{A}}_m =  
\breve{A}_m \ , \ \hat{x}_m = -i \hbar X_{x^i} =
i \hbar \displaystyle{\frac{\partial}{\partial \breve{A}_m}}}\}$ or  
by ${\{ \hat{\breve{A}}_m = -i \hbar X_{\breve{A}_m} = -i  
\hbar\displaystyle{\frac{\partial}{\partial x^m}} \ , \ \hat{x}^m =  
x^m }\}$, respectively.

In both representation the commutator between the quatum operators  
related to these representations is given by $(-i \hbar)$.

\begin{equation}
[ e \hat{\breve{A}_m} \ , \ \hat{x^n} ] \Psi = -i \hbar \delta_m ^n \Psi
\end{equation}
\label{six}

Thus, for the relevant direction in flux quantization $\oint e A_l  
dl$ one obtains:

\begin{equation}
[ e \hat{\breve{A}_l} \ , \ \hat{l} ] = -i \hbar
\end{equation}
\label{seven}

Equivalently, we have according to the general quantum mechanics a  
true uncertainty relation for $\breve{A}_l$ and $l$, i. e.: $e  
\Delta \breve{A}_l \cdot \Delta l \sim \hbar$ or $e \Delta  
\breve{A}_l \cdot \Delta l \geq \hbar$. In other words, to  
understand the flux quantization and to describe it according to the  
canonical quantization sheme, one has to consider the operator  
commutator (6) or (7) and equivalently the related uncertainty  
relation. Thus, we derived the new uncertainty relation within a  
consistent quantization formalism.

\bigskip

Accordingly, in view of the fact that $\Delta l \geq l_B$ in  
quantum electrodynamics under strong magnetic fields, a pure gauge  
potential should have in view of $e \Delta \breve{A}_l \cdot \Delta  
l \geq \hbar$ a maximal uncertainty of $(\Delta \breve{A}_l)_{max} =  
\displaystyle{\frac{\hbar}{e l_B}}$. In other words, the pure edge  
potential $A$ which must exists calssically only exactly on the  
edges of the QHE sample [2] and must be zero on the sample, is  
quantum electrodynamically however not zero on the sample: But it  
has for the $A := \breve{A}_l$ according to the uncertainty relation  
$e \Delta \breve{A} \cdot l_B \sim \hbar$ a non-vanishing value  
("the quantum potential drop") over the edge of sample.

\medskip
On the other hand, in view of the relations between the magnetic  
field strength $B$, magnetic length and the global density of  
electrons $n$ with the filling factor $\nu$ in QHE, i. e. $l^2_B =  
{\displaystyle{\frac{\hbar}{eB}}} = {\displaystyle{\frac{\nu}{2\pi  
n}}}$, it is obvious that a variation of only one of these factors  
changes the magnetic length and so it changes also the current  
position and the potential distribution on the sample. However, if  
$B$ or ${\displaystyle{\frac{\nu}{n}}}$ remain the same for a set of  
IQHE samples in an experiment, then the magnetic length and so also  
the potential uncertainty should be invariant for all these samples  
under the IQHE conditions independent of their geometries and other  
factors.

These are the quantum theoretical basics of what is observed in the  
mentioned experiments for the potential drops for two different  
sets of samples with two different filling factors as in \cite{DK}  
and \cite{font} \cite{more}. In the case \cite{DK} the authors  
report on the observation of potential drops across the IQHE-samples  
over a length of $100 \mu m$ from the edge of samples. We show that  
this potential drop which has the {\it magnitude} of  
$(l_B^{-1})_{[6]}$ for the QH-sample used in Ref. \cite{DK} is the  
same as the uncertainty for potential $(\Delta \breve{A}_l)_{max}$  
given by the uncertainty relation $ e (\Delta \breve{A}_{max})_{[7]}  
\cdot (l_B)_{[7]} = \hbar$ \cite {nn}, [9]. Thus, we identify the  
{\it maximal quantum} potential drop in QHE with the quantum  
electrodynamical uncertainty for the value of potential $\Delta  
\breve{A}_{max} = \displaystyle{\frac{\hbar}{e l_B}}$ on the edge of  
each sample for the given $l_B$ according to the  
${\displaystyle{\frac{\nu}{n}}}$ value of the same sample.

Recall however, that according to the uncertainty relation for  
$\Delta l > l_B$ one should have $\Delta \breve{A} <  
\breve{A}_{max}$. In other words, for the "ideal" case where the  
electronic current flow within the $l_B$ width over the edge of QHE  
sample \cite{kk}, the potential drop has its maximal value $\Delta  
A_{max}$. This is variefied in experiments [7] and [8]. But, if in  
QHE the current flow further within the sample, then the potential  
drop $\Delta A$ is less than its maximal value. This is variefied in  
some of experiments in Ref. [9].

\medskip
Furthermore, as we mentioned above the electromagnetic potential is  
in view of its gauge dependence non-observable. The observables  
related with the potential or those related with its field strength  
are phase angles given by the closed path integral of potential or  
the surface integral of field strength, which are observable by the  
quantum mechanical interfrence patterns [3]. Equivalently, a  
constant potential multiplied by a proper length, e. g. by the  
circumference of mentioned closed path is also observable. For  
example according to the definition of magnetic length $l^{2}_B =  
{\displaystyle{\frac{\hbar}{e B}}}$ we have (see also below):

\begin{equation}
(l_B)^2 \cdot B = l_B \cdot \Delta \breve{A}_{max} =  
\frac{\hbar}{e}\;\;\;,
\end{equation}
\label{eight}

which is equivalent to the definition of magnetic flux quantum  
through $\int\int B ds = \oint \breve{A}_m dx^m = N  
{\displaystyle{\frac{h}{e}}}$ for $N = 1$ case. Another observable  
of potential is the difference of potential or in quantum case the  
uncertainty $\Delta A = A - {\AA}$.

Moreover, let us mention that from the uncertainty relation

\begin{equation}
e \Delta \breve{A}_l \cdot \Delta l \sim e B\cdot \Delta x^m \cdot  
\Delta x^n \epsilon_{mn} \sim \hbar
\end{equation}
\label{nine}

it is obvious that the in the geometric units where  
$\displaystyle{\frac{\hbar}{e}}$ is considered as dimensionless the  
potential uncertainty is given in $(Tesla \cdot L)$. In other words,  
the Potential uncertainty can be observed, in view of $L^{-2}$  
dimension of the magnetic field strength, either in $L^{-1}$ or it  
can appear with respect to a fixed magnetic field value in $L$.

\medskip
Therefore, if one considers the quantum electrodynamical  
uncertainty relation $\Delta \breve{A}_{max} \cdot l_B =  
{\displaystyle{\frac{\hbar}{e}}}$, then, one obtains with the given  
$l_B$ according to the data in Ref. \cite{DK} for $ (\Delta  
\breve{A}_{max})_{[7]} = {\displaystyle{\frac{\hbar}{e}}}\cdot  
(l^{-1}_B)_{[7]}$ a value about $100 \mu m$ for $(\Delta  
\breve{A}_{max})_{[7]}$, which is the mentioned observed width for  
potential drops \cite{DK} \cite{dim}.
This result show that in view of the definition of magnetic length  
the measured value of potential drops is a fundamental value for the  
given $l_B$ value of each IQHE sample independent of other  
parameters of that sample. Thus, the quantum potential drop on the  
edge of QH-samples is nothing than the uncertainty of potential or  
the width where in view of quantum situation the pure gauge  
potential exists and does not vanish although it should vanish there  
classically.

Therefore, practically what is measured in \cite{DK} and  
\cite{font} is the uncertainty relation $\Delta \breve{A}_{max}  
\cdot l_B = {\displaystyle{\frac{\hbar}{e}}}$ for the samples under  
consideration which are represented by their own $l_B$ values given  
according to their own $\displaystyle{\frac{\hbar}{eB}}$ or  
$\displaystyle{\frac{\nu}{2\pi n}}$ values. Accordingly, in view of  
the fact that $l_B$ is implemented in - and given by - the sample  
quantities and also that $\displaystyle{\frac{\hbar}{e}}$ is a  
universal constant, one observes in such experiments within what  
width from the edges the $\Delta A$ is still existing quantum  
theoretically.

\medskip
The same calculation can be done for the experiments with filling  
factor ${\nu}^{\prime} = 4$ about which it is reported in Ref. \cite  
{font}. The theoretical result agrees also in this case with the  
measured result.

\medskip
To be precize, let us mention that in other experiments  
\cite{font}, where the electronic concentration is almost the {\it  
same} as in Ref. \cite{DK} but the filling factor is ${\nu}_{[8]}  
=4$, one observed potential drops of $\approx 70 \mu m$. This is in  
good agreement with our theoretical result, since for ${\nu}_{[8]}  
=4$ one obtains according to the data of Ref. \cite{font} a magnetic  
length $(l_B)_{[8]} \approx 1.4 \cdot 10^{-2}$ $\mu m$. Thus, the  
theoretical value of $( \Delta \breve{A}_{max})_{[8]} =  
{\displaystyle{\frac{\hbar}{e}}}(l_B^{-1})_{[8]}$ becomes $\approx  
70 \mu m$ which is indeed the measured value according to Ref.  
\cite{font} (see also \cite{dim}).

\medskip
Therefore, one should claim that the measured potential drop or  
penetration length of electromagnetic potential on the edge of every  
QHE sample should depend, according to the theoretical value of $  
\Delta \breve{A} = {\displaystyle{\frac{\hbar}{e}}}(l_B)^{-1}$, only  
on the $l^{-1}_B$ value of the sample under consideration  
\cite{dim}.

\medskip
conversely, the fact that the ratio between potential drops in  
\cite{DK} and \cite{font} with the same density $n$ is given by a  
factor of $\displaystyle{\frac{(\Delta  
\breve{A}_{max})_{[7]}}{(\Delta \breve{A}_{max})_{[8]}}} = 1.4$ and  
this is equal to the ratio  
$(\displaystyle{\frac{{\nu}_{[8]}}{\nu_{[7]}}})^{\frac{1}{2}} =  
(\displaystyle{\frac{4}{2}})^{\frac{1}{2}}$ and further that $l_B =  
(\displaystyle{\frac{\nu}{2\pi n}})^{\frac{1}{2}}$ for every QHE
sample, manifests the fact that potential drop for each sample must  
be a function of its own $l_B$ only. Otherwise, the mentioned ratio  
can not be such a simple number and one should see an other ratio  
by comparision of potential drops results of two groups of  
experiments \cite{DK} and \cite{font} .

\medskip
Furthermore, it is expected that the observed length of the  
potential drop should be related with parameters of samples. This is  
indeed true in our model, if one recalls that here the potential  
drop is given by the reciproc of  magnetic length and this one is  
given by the concentration of charge carriers which is indeed the  
main parameter of a sample.

\medskip
In conclusion let us mention that such a penetration length is also  
comparable with London's penetration length in superconductivity  
\cite{sup}.

\bigskip
Footnotes and references


\begin{thebibliography}{100}
\bibitem{almost}
A pure gauge potential $\breve{A}_m$ with $F_{mn} (\breve{A}_m) =  
0$ in QHE case is a potential in a multiply connected region, thus  
it can not be gauged away everywhere in the region. Since, the  
general field strength is in some parts of the region zero and in  
some other parts not. Quantum mechanically one can observe the  
influence of such pure gauge potential also in those regions where  
the field strength is zero (see Bohm-Aharonov effect). Recall also  
that the configuration space of n-particles in two spatial  
dimensions is non-simply connected.
\bibitem{current}

It is intresting to mention also that the integration path in the  
usual derivation of flux quantization is chosen in a manner that the  
electronic current becomes zero on this path. In view of the fact  
that, the vanishing of current in some regions is equivalent to the  
definition of the electromagnetic pure gauge potential in these  
regions, the electromagnetic potential on the integration path is a  
pure gauge potential (see Ref.\cite{almost}) which results, as it is  
shown above, in the flux quantization.
\bibitem{comp}
This procedure of canonical comparison to identify the true phase  
space variables of a new system for its quantization is the very  
general and canonical one which is used for any quantization, e. g.  
for the quantization of the Yang-Mills action functional as like as  
Maxwell's electrodynamics, Chern-Simons action functional, etc. See  
for the Chern-Simons example Ref. [2].
\bibitem{clq}
One main reason for this procedure of quantization is that  
classical ( symplectic ) mechanics should be the classical limit of  
quantum mechanics. Therefore, the quantum operation of vector fields  
should results in classical limit in the classical operation of  
them.
\bibitem{nn}
According to the data about the IQHE samples in Ref. \cite{DK} the  
global concentration is

$n = 3.7 \cdot 10^{11} cm^{-2}$ and $\nu = 2$. Thus, one obtains  
$l_B \approx 10^{-2} \mu m$ for the samples used in Ref. \cite{DK}.

The measured pentration length is given to be about $100 \mu m$  
which is almost exactly $|l^{-1}_B| \mu m$.
\bibitem{sup}
It is well known that superconducting effects can be considered as  
to be related with the QHE: see R. B. Laughlin in Ref. [1d].
\end{thebibliography}
\end{document}